\documentstyle[twoside,amssymb,12pt]{article}

\newcommand{\nc}{\newcommand}
\nc{\la}{\lambda} \nc{\alf}{\alpha}
\nc{\tht}{\theta}  \nc{\be}{\beta}  \nc{\eps}{\epsilon}
\nc{\ga}{\gamma}  \nc{\D}{\Delta}  \nc{\G}{\Gamma}  \nc{\vphi}{\varphi}
\nc{\de}{\delta} \nc{\si}{\sigma}  \nc{\ka}{\kappa}   \nc{\Si}{\Sigma}
\nc{\om}{\omega}  \nc{\Om}{\Omega}  \nc{\z}{\zeta}
\nc{\qq}{\quad\quad}   \nc{\nf}{\infty}   \nc{\pt}{\partial}
\nc{\dl}{\mathop{\smash{\cal L}}}  \nc{\black}{\rule{1mm}{3mm}}
\nc{\beq}{\begin{equation}}     \nc{\eeq}{\end{equation}}
\nc{\beqa}{\begin{eqnarray}}  \nc{\dst}{\displaystyle}  \nc{\sst}{\scriptstyle}
\nc{\eeqa}{\end{eqnarray}} \nc{\nnb}{\nonumber}
\nc{\bs}{\backslash}        \nc{\mb}{\mathbb}
\nc{\sn}{{\rm sn}\,} \nc{\cn}{{\rm cn}\,}     \nc{\dn}{{\rm dn}\,}
\nc{\ti}{\tilde}         \nc{\wti}{\widetilde}  \nc{\wh}{\widehat}
\nc{\ol}{\overline}      \nc{\ul}{\underline}

\newcounter{muni}
\newenvironment{remunerate}{\begin{list}{{\rm \arabic{muni}.}}
{\usecounter{muni}
\setlength{\leftmargin}{0pt}\setlength{\itemindent}{38pt}}}{\end{list}}

\nc{\brm}{\begin{remunerate}}   \nc{\erm}{\end{remunerate}}

\newtheorem{nth}{Proposition}  \newtheorem{nlem}{Lemma}
\nc{\simas}{\mathop{\smash{\sim}}}
\nc{\stg}{\mathop{\smash{*}}}   \nc{\LIM}{\mathop{\smash{\lim}}}
\nc{\st}{\mathop{\smash{\delta}}} \nc{\SUP}{\mathop{\smash{\rm sup}}}
\nc{\barr}{\begin{array}}   \nc{\earr}{\end{array}}   \nc{\dg}{\dagger}
\nc{\mtvb}{\mathversion{bold}}   \nc{\mtvn}{\mathversion{normal}}

\begin{document}
\begin{titlepage}

\vskip 1.0truecm 
\centerline{\Large\bf Some cubic birth and death processes}

\vspace{5mm}
\centerline{\Large\bf and their related orthogonal polynomials}

\vspace*{1cm}
\centerline{\bf Jacek GILEWICZ${}^{\;*}$, Elie LEOPOLD${}^{\;*}$,}

\vspace{5mm}
\centerline{\bf Andreas RUFFING${}^{\;\dagger}$, Galliano VALENT${}^{\;\ddagger}$}

\vspace{5mm}
\centerline{${}^{*}$\it CNRS Luminy Case 907 }
\centerline{\it Centre de Physique Th\'eorique}
\centerline{\it F-13288 Marseille Cedex 9, France}
\vskip 0.3truecm
\centerline{\it ${}^{\dagger}$ Zentrum Mathematik}
\centerline{\it Technische Universit\" at M\" unchen}
\centerline{\it Boltzmannstra{\ss}e 3, D-85747 Garching, Germany}
\vskip 0.3truecm
\centerline{${}^{\ddagger}$ \it Laboratoire de Physique Th\'eorique et des
Hautes Energies}
\centerline{\it CNRS, Unit\'e associ\'ee URA 280}
\centerline{\it 2 Place Jussieu, F-75251 Paris Cedex 05, France}
\nopagebreak
\vskip 0.5truecm

\begin{abstract}
\noindent The orthogonal polynomials with recurrence relation
\[(\la_n+\mu_n-z)\,F_n(z)=\mu_{n+1}\,F_{n+1}(z)+\la_{n-1}\,F_{n-1}(z)\]
with two kinds of cubic transition rates $\la_n$ and $\mu_n,$  
corresponding to indeterminate Stieltjes
moment problems, are analyzed. We derive generating functions for these two 
classes of polynomials, which enable us to compute their Nevanlinna matrices. We 
discuss the asymptotics of the Nevanlinna matrices in the complex plane.
\end{abstract}
%\vfill {\it E-Mail:}\\
%{\it 1) gilewicz@cpt.univ-mrs.fr}\\
%{\it 2) leopold@cpt.univ-mrs.fr}\\
%{\it 3) valent@cpt.univ-mrs.fr}
\end{titlepage}

\newpage\nc{\cp}{{\cal P}}
\section{Introduction}
The field of indeterminate moment problems applied to birth and death processes 
has been quite active in the past ten years and many explicit examples have 
been worked out, 
see \cite{BV} and the many references therein. Restricting ourselves to the
case of polynomial transition rates $\la_n$ and $\mu_n$ the results obtained 
dealt mainly with quartic rates \cite{BV},\cite{Va1}. It is the aim of this article
to show that the same underlying ideas that led successfully to the computation 
of the Nevanlinna matrices for the quartic rates can be applied to some cubic
rates, leading to some explicit integral representations for their Nevanlinna 
matrix elements.  

The plan of the article is the following. In Section 2 we will recall some 
basic relations between birth and death processes and orthogonal polynomial theory. 
In this article we will be concerned with the two processes:
\[ (P1):\quad\la_n=(3n+3c+1)^2(3n+3c+2), \qq\mu_n=(3n+3c-1)(3n+3c)^2(1-\de_{n0}),\]
and
\[ (P2):\quad\la_n=(3n+3c+1)(3n+3c+2)^2, \qq\mu_n=(3n+3c)^2(3n+3c+1)(1-\de_{n0}),\]
under the assumption $c>0.$ In Section 3 some background material useful in 
the sequel is gathered. In Section
4 and 5 we obtain some generating functions which will allow, in Section 6 to 
compute the Nevanlinna matrices for both processes. In Section 7 we analyze the
asymptotics, in the complex plane, of the Nevanlinna matrix elements.

\section{Birth and death processes versus orthogonal polynomials}
Birth and death processes are special stationary Markov processes whose state
space is ${\mb N}$, representing for instance some population.
We are interested in the time evolution of such a population, described by the
transition probabilities $\cp_{m,n}(t)$ yielding the probability that the population
goes from the state $m$ at time $t=0$ to the state $n$ at time $t>0.$ This evolution
is supposed to be governed by
\[\barr{l}
\cp_{n,n+1}(t)=\la_n\,t+o(t),\\[4mm]
\cp_{n,n}(t)=1-(\la_n+\mu_n)t+o(t),\\[4mm]
\cp_{n,n-1}(t)=\mu_n\,t+o(t),\earr\qq t\to 0.\]
For applications the most important problem is to find $\cp_{m,n}(t)$
for given rates $\la_n$ and $\mu_n,$ with suitable extra constraints to be
described later on.

\ From the previous setting one can prove that the transition probabilities 
have to be a solution of the forward Kolmogorov equations
\beq
\frac{d}{dt}\cp_{m,n}=\la_{n-1}\,\cp_{m,n-1}+\mu_{n+1}\,\cp_{m,n+1}
-(\la_n+\mu_n)\,\cp_{m,n}.\eeq
The $\cp_{m,n}(t)$ are assumed to be continuous for small time scales with
\beq\label{op2}
\lim_{t\to 0}\ \cp_{m,n}(t)=\de_{m,n}.\eeq
A representation theorem for $\cp_{m,n}(t)$ was proved by Karlin and MacGregor in
\cite{kmg} which links birth and death processes and orthogonal polynomials theory.
Let us define the polynomials $F_n(x)$ by the three-terms recurrence relation
\beq\label{op1}
(\la_n+\mu_n-x)F_n(x)=\mu_{n+1}F_{n+1}(x)+\la_{n-1}F_{n-1}(x),\qq n\geq 1,\eeq
with the initial conditions
\[F_{0}(x)=1,\qq\quad F_1(x)=\frac{\la_0+\mu_0-x}{\mu_1}.\]
Let us define
\[\pi_0=1,\qq\qq\pi_n=\frac{\la_0\la_1\cdots\la_{n-1}}{\mu_1\mu_2\cdots\mu_n},
\qq n\geq 1.\]
If the positivity conditions
\beq
\la_n>0,\qq n\geq 0,\qq\mbox{and}\qq \mu_0=0,\qq \mu_n>0,\qq n\geq 1\eeq
are fulfilled, then there is a positive measure $\psi$ for which
\beq
\cp_{m,n}(t,\psi)= \frac 1{\pi_m}\int_{{\rm supp}(\psi)}\ e^{-xt}F_m(x)F_n(x)\,d\psi(x).
\eeq
Then the initial condition (\ref{op2}) is nothing but the orthogonality relation
\[\frac 1{\pi_m} \int_{{\rm supp}(\psi)}\ F_m(x)F_n(x)\,d\psi(x)=\de_{m,n}.\]
Such a measure has well-defined moments
\[c_n=\int_{{\rm supp}(\psi)}\ x^n\,d\psi(x),\qq\qq n=0,\,1,\ldots.\]
If ${\rm supp}(\psi)\subseteq{\mb R}$ this is a Hamburger moment problem and if
${\rm supp}(\psi)\subseteq [0,+\nf[$ this is a Stieltjes moment problem. In the event
that the measure $\psi$ is not unique we speak of indeterminate Hamburger (or
indeterminate Stieltjes) moment problems, indet H or indet S for short. Stieltjes
(see \cite{Ak}) obtained the necessary and sufficient conditions for a
moment problem to be indet S
\beq\label{indets}
\sum_{n\geq 0}\,\pi_n<\nf,\qq\sum_{n\geq 1}\,\frac 1{\la_n\pi_n}<\nf.\eeq
These conditions imply that it is also indet H. 

\noindent Let us consider now the two processes to be analyzed in this article. We 
will denote the first one as the process $P1$, with rates
\beq\label{rate1}
\la_n=(3n+3c+1)^2(3n+3c+2), \qq\mu_n=(3n+3c-1)(3n+3c)^2(1-\de_{n0}),
\quad n\geq 0,\eeq
and the second one as process $P2$, with rates
\beq\label{rate2}
\la_n=(3n+3c+1)(3n+3c+2)^2,  \qq\mu_n=(3n+3c)^2(3n+3c+1)(1-\de_{n0}),
\quad n\geq 0\eeq
Using the notation $\dst\,(a)_n=\G(n+a)/\G(a)\,$ we have for $P1$ the 
large $n$ behaviour
\[\pi_n=\left(\frac{(c+1/3)_n}{(c+1)_n}\right)^2={\cal O}(n^{-4/3}),
\qq\qq \frac 1{\pi_n\,\mu_n}={\cal O}(n^{-5/3})\]
and for $P2$
\[\pi_n=\frac{(c+1/3)_n\,((c+2/3)_n)^2}{((c+1)_n)^2\,(c+4/3)_n}={\cal O}(n^{-5/3}),
\qq\qq \frac 1{\pi_n\,\mu_n}={\cal O}(n^{-4/3})\]
These asymptotic estimates show that the conditions (\ref{indets}) are satisfied 
and therefore the two processes are indet S and indet H.

\section{Background material}
In order to describe the Nevanlinna matrix we will need a triplet of elementary
functions defined by
\beq\label{b1}
\si_l(u)=\sum_{n\geq 0}(-1)^n\,\frac{u^{3n+l}}{(3n+l)!},\qq\quad l=0,1,2.\eeq
It is easy to check the relations
\beq\label{b2}\barr{lll}
\si_1'=\si_0,\qq & \si_2'=\si_1,\qq& \si_0'=-\si_2,\\[4mm]
\si_0(0)=1,\qq & \si_1(0)=0,\qq & \si_2(0)=0.\earr\eeq
These functions are called trigonometric functions of order 3, since they are
three linearly independent solutions of the third order differential equation
\[\si_l'''+\si_l=0,\qq l=0,\,1,\,2.\]
Their explicit form is
\beq\label{sigma}\barr{l}\dst
\si_0(u)=\frac 13(e^{-u}+e^{j u}+e^{\ol{j}u})=
\frac 13\left(e^{-u}+2\cos\left( \frac{\sqrt{3}}{2}u\right)e^{u/2}\right),
\\[4mm]\dst
\si_1(u)=\frac 13(-e^{-u}+\ol{j}e^{j u}+j e^{\ol{j}u})=\frac 13
\left(-e^{-u}+2\cos\left(\frac{\sqrt{3}}{2}u-\frac{\pi}{3}\right)e^{u/2}\right),
\\[4mm]\dst
\si_2(u)=\frac 13(e^{-u}-j e^{j u}-\ol{j}e^{\ol{j} u})=\frac 13\left(e^{-u}
-2\cos\left(\frac{\sqrt{3}}{2}u+\frac{\pi}{3}\right)e^{u/2}\right),\earr
\quad j=e^{i\pi/3}.\eeq

We will need also the following functions
\beq\label{fcts}
\tht(t)=\int_0^t\,\frac{du}{(1-u^3)^{2/3}},\qq
\wh{\tht}(t)=\tht_0-\tht(t),\qq 
\tht_0\equiv\int_0^1\,\frac{du}{(1-u^3)^{2/3}}=\frac{\G^3(1/3)}{2\pi\sqrt{3}}.\eeq
Observing that $\,\wh{\tht}(t)\,$ is continuous, decreasing and concave for 
$\,t\in\,[0,1]$ gives the bounds 
\beq\label{encadre}
1-t\leq\frac{\wh{\tht}(t)}{\tht_0}\leq 1.\eeq

\section{First generating function}
We will consider, for $c>0,$ the slightly more general rates than the ones 
defined in (\ref{rate1}):
\beq\label{gf1}\barr{lll}
\la_n=(3n+3c+1)^2(3n+3c+2),\qq & n\geq 0 & \\[4mm]\dst 
\mu_n=(3n+3c-1)(3n+3c)^2,\qq & n\geq 1 &\qq \mu_0\geq 0,\earr\eeq
where $\mu_0$ is taken as a free parameter, not necessarily equal 
to $(3c-1)(3c)^2.$

We will denote by $F_n(z;c,\mu_0)$ the polynomials with recurrence relation 
(\ref{op1}) and the rates (\ref{gf1}). Obviously the polynomials corresponding to 
process $P1$ are recovered as the limiting values $F_n(z;c,0).$

To get most conveniently a generating function for them, it is useful to define 
a triplet of functions $d_{3n+l}(\z),\ l=0,1,2$  for $n=0,1,\ldots$ by 
the recurrence relation
\beq\label{gf2}
n\geq 0 \qq 
\left\{\barr{ll}
d_{3n+1}=-\z\,d_{3n}+\mu_n\,d_{3n-2}, & \qq\qq (a)\\[4mm]  
d_{3n+2}=-\z\,d_{3n+1}, & \qq\qq (b)\\[4mm] 
d_{3n+3}=-\z\,d_{3n+2}+\la_n\,d_{3n}, & \qq\qq (c)\earr\right.\eeq
with the initial values
\[d_{-2}=\frac 1{\z^2},\qq\qq d_0=1,\]
and the definition $\,\z=z^{1/3}.$ Note that $d_{3n}(\z)$ are polynomials with 
respect to $z.$\\[0.2cm]

\noindent Let us begin with
\begin{nth}\label{rel1} The polynomials $F_n$ and $d_n$ are related by
\beq\label{gf3}
F_n(z;c,\mu_0)=(3c)!\ \frac{(c+1/3)_n}{(c+1)_n}\,\frac{d_{3n}(\z)}{(3n+3c)!},\eeq
where we use the notation $\,(\alf)!=\G(\alf+1)\,$ for $\,\alf>0.$
\end{nth}

\noindent{\bf Proof :}

\noindent Let us define $\,M_n(z)=d_{3n}(\z).$ Using (\ref{gf2}c) and  (\ref{gf2}b) 
in (\ref{gf2}) we have for $n\geq 1$
\[M_{n+1}=d_{3n+3}=-\z\,d_{3n+2}+\la_n\,d_{3n}=\la_n\,M_n+\z ^2\,d_{3n+1}=
(\la_n-z)M_n+\z ^2(\z\,d_{3n}+d_{3n+1}).\]
Then we use (\ref{gf2}a) and (\ref{gf2}b) with the shift $n\to n-1$  
to get
\[M_{n+1}=(\la_n-z)M_n+\mu_n\,\z ^2\,d_{3n-2}=(\la_n-z)M_n-\mu_n\,\z\,d_{3n-1}.\]
The term involving $d_{3n-1}$ is disposed of using the (\ref{gf2}c) with the 
shift $n\to n-1.$ One is left with
\[M_{n+1}=(\la_n-z)M_n+\mu_n(d_{3n}-\la_{n-1}\,d_{3n-3})=(
\la_n+\mu_n-z)M_n-\la_{n-1}\,\mu_n\,M_{n-1},\qq\quad n\geq 1.\]
The boundary conditions are to be computed separately and give
\[M_0=1,\qq\qq M_1=d_3=\la_0+\mu_0-z.\]
It is then easy to check that the polynomials $F_n$ are related to the $M_n$ by
\beq\label{gf4}
F_n(z;c,\mu_0)=\frac{M_n(z)}{\mu_1\mu_2\cdots\mu_n}=
\frac{d_{3n}(\z)}{\mu_1\mu_2\cdots\mu_n},\qq n\geq 0.\eeq
Using Gauss multiplication formula we have
\[\frac{(3n+3c)!}{(3c)|}=3^{3n}(c+1/3)_n(c+2/3)_n(c+1)_n=
\frac{(c+1/3)_n}{(c+1)_n}\,\mu_1\mu_2\cdots\mu_n,\]
and this leads to the desired relation (\ref{gf3}).$\qq\Box$

In view of Proposition \ref{rel1} we need generating functions for 
$d_{3n+l}$ which we define for the variable $t$ -- in a suitable neighbourhood of 
the origin --  as
\beq\label{gf5}
G_l(\z,t)=\sum_{n\geq 0}\,d_{3n+l}(\z)\,\frac{t^{3n+3c+l}}{(3n+3c+l)!},\qq l=0,1,2.
\eeq
Routine computations, using relations (\ref{gf2}) give for these generating functions 
the linear differential system
\beq\label{gf6}
\barr{rcl} 
(1-t^3)\,D_t\,G_0-t^2\,G_0+\z\,G_2 & = &\dst \frac{t^{3c-1}}{(3c-1)!},\\[5mm] 
(1-t^3)\,D_t\,G_1-2t^2\,G_1+\z\,G_0 & = &\dst \frac{\mu_0}{\z^2}\,\frac{t^{3c}}{(3c)!},
\\[5mm]
D_t\,G_2+\z\,G_1 & = & 0.\earr\eeq
All the factorials involved are well defined in view of the 
hypothesis $\,c>0.$

Switching from the functions $G_i$ to new functions $\wh{G}_i$ defined by 
\beq\label{gf7}
G_0=\frac{\wh{G}_0}{(1-t^3)^{1/3}},\qq G_1=\frac{\wh{G}_1}{(1-t^3)^{2/3}},
\qq G_2=\wh{G}_2,\eeq
the differential system takes the more symmetric form 
\beq\label{gf8}
\barr{lcl} 
(1-t^3)^{2/3}\,D_t\,\wh{G}_0+\z\,\wh{G}_2 & = & \dst\frac{t^{3c-1}}{(3c-1)!},\\[5mm]
(1-t^3)^{2/3}\,D_t\,\wh{G}_1+\z\,\wh{G}_0 & = & \dst\frac{\mu_0}{\z^2}\,
\frac{t^{3c}}{(3c)!}\,(1-t^3)^{1/3},\\[5mm]
(1-t^3)^{2/3}\,D_t\,\wh{G}_2+\z\,\wh{G}_1 & = & 0.\earr\eeq
Using the variable $\tht(t)$ defined in (\ref{fcts}) we observe that 
$\,(1-t^3)^{2/3}\,D_t=D_{\tht}\,$ so that (\ref{gf8}) becomes an inhomogeneous 
differential system with constant coefficients:
\beq\label{gf8bis}
D_{\tht}\,\wh{G}_0+\z\,\wh{G}_2=a(\tht),\qq 
D_{\tht}\,\wh{G}_1+\z\,\wh{G}_0=\frac{b(\tht)}{\z^2},\qq 
D_{\tht}\,\wh{G}_2+\z\,\wh{G}_1=0.\eeq
This is easily solved for $\wh{G}_0$; one gets
\beq\label{gf8ter}
\wh{G}_0(\tht)=\int_0^{\tht}\,\si_0(\z(\tht-v))\,a(v)\,dv+\int_0^{\tht}\,
\frac{\si_2(\z(\tht-v))}{\z^2}\,b(v)\,dv.\eeq
From this result we recover $\wh{G}_0(z,t)$ by coming back to the original variable $t$ 
and after the change of variable $v=\tht^{-1}(u)$ in the integral. Using the notation 
$\Theta(t,u)=\tht(t)-\tht(u),$ we conclude to:
\beq\label{gfquater}
\barr{l}\dst 
\wh{G}_0(z,t)=\int_0^t\si_0(\z\Theta(t,u))\frac{u^{3c-1}}{(3c-1)!}\,(1-u^3)^{-2/3}\,du
\\[4mm] \dst \hspace{7cm} +\mu_0
\int_0^t\frac{\si_2(\z\Theta(t,u))}{\z^2}\frac{u^{3c}}{(3c)!}\,(1-u^3)^{-1/3}\,du.
\earr
\eeq
Taking into account relations (\ref{gf7}), (\ref{gf4}) and (\ref{gf3}) we have on 
the one hand
\[\wh{G}_0(z,t)=(1-t^3)^{1/3}\,G_0(z,t)=
\frac 1{(3c)!}\sum_{n\geq 0}\frac{(c+1)_n}{(c+1/3)_n}\,F_n(z;c,\mu_0)\,t^{3n+3c}
\,(1-t^3)^{1/3},\]
and on the other hand $\wh{G}_0(z,t)$ given by (\ref{gfquater}). Gathering all 
these pieces we end up with

\begin{nth} The polynomials $F_n(z;c,\mu_0)$ have the generating function
\beq\label{gftop1}\barr{l}\dst 
\sum_{n\geq 0}\frac{(c+1)_n}{(c+1/3)_n}\,F_n(z;c,\mu_0)\,t^{3n+3c}
\,(1-t^3)^{1/3}=
\\[6mm]\dst 
3c\,\int_0^t\,\si_0(\z\Theta(t,u))\,u^{3c-1}(1-u^3)^{-2/3}\,du+
\mu_0\,\int_0^t\frac{\si_2(\z\Theta(t,u))}{\z^2}\,u^{3c}(1-u^3)^{-1/3}\,du.
\earr\eeq
\end{nth}
This is not quite enough to compute the Nevanlinna matrix; in fact we need
\beq\label{ff}
{\cal F}(z;c,\mu_0)=\sum_{n\geq 0}\,F_n(z;c,\mu_0).\eeq 
Using the notation
\[B(\alf,\be)=\frac{\G(\alf)\G(\be)}{\G(\alf+\be)},\]
we will now state:

\begin{nth} We have the relations
\beq\label{gftop2}\barr{l}\dst
{\cal F}(z;c,\mu_0)=
\frac{3}{B(c+1/3,2/3)}\left\{3c\int_0^1\,u^{3c-1}(1-u^3)^{-2/3}\,
\frac{\si_1(\z\wh\tht(u))}{\z}\,du\right.\\[6mm]\dst \left.\hspace{6.5cm}
+\mu_0\int_0^1\,u^{3c}(1-u^3)^{-1/3}\,\frac{1-\si_0(\z\wh\tht(u))}{z}\,du
\right\},\earr\eeq
valid for $c>0$ and
\beq\label{gftop3}\barr{l}\dst 
1-\frac z{\mu_0}\,{\cal F}(z;c,\mu_0)=\frac 3{B(c-2/3,2/3)}\,
\frac{(3c-1)(3c)^2}{\mu_0}\int_0^1\,
u^{3c-3}(1-u^3)^{-1/3}\,\si_0(\z\wh\tht(u))\,du\\[6mm]\dst 
\hspace{1cm}+\frac{[\mu_0-(3c-1)(3c)^2]}{\mu_0}\frac{3}{B(c+1/3,2/3)}\,
\int_0^1\,u^{3c}(1-u^3)^{-1/3}\si_0(\z\wh\tht(u))\,du,\earr\eeq
valid for $c>1.$
\end{nth}

\noindent{\bf Proof :}
In (\ref{gftop1}) we set $t=\tau^{1/3},$ then we multiply both sides by 
$\,\tau^{-2/3}(1-\tau)^{-2/3}\,$ and integrate from $\tau=0$ to $\tau=1.$ 
The left hand side integral involves a Eulerian integral and we get
\[\frac{\G(c+1/3)\G(2/3)}{\G(c+1)}\,\sum_{n\geq 0}\,F_n(z;c,\mu_0)=
3(3c)!\int_0^1\frac{\wh{G}_0(z,t)}{(1-t^3)^{2/3}}\,dt.\]
The right hand side is a double integral, which, upon interchange of the order of 
the integrations and use of relations (\ref{b2}), gives (\ref{gftop2}). In this 
last result, the integral with no $\si$ function, when expressed in terms of Euler 
gamma functions simplifies to $\mu_0/z.$ The first integral in (\ref{gftop2}), 
using relations (\ref{b2}), can be integrated by parts twice; then 
elementary algebra results in (\ref{gftop3}).  $\qq\Box$

The results obtained so far are sufficient to compute the functions $C_1(z)$ and 
$D_1(z)$ in the Nevanlinna matrix of process $P1.$ However, to get the full matrix 
we need also 
the generating function for the dual process in the sense of Karlin and McGregor 
(KMG for short). For the reader's convenience let us recall its definition.

The correspondence from a process ${\cal P}$ to its KMG dual $\wti{\cal P}$ 
is as follows
\[{\cal P}=\{\la_n,\,\mu_n\}\quad\to\quad 
\wti{\cal P}=\{\wti{\la}_n=\mu_{n+1},\,\wti{\mu}_n=\la_n\}.\]
It follows that the dual process of $P1$  will have 
\beq\label{gf9}
\wti{\la}_n=(3n+3c+2)(3n+3c+3)^2,\qq\wti{\mu}_n=(3n+3c+1)^2(3n+3c+2),\quad n\geq 0,
\eeq
which correspond to the process $P2$ up to the shift $c\to c+1/3.$ So we will now 
work out a generating function for the process $P2.$

\section{Second generating function}
Here again we will consider, for $c>0,$ the slightly more general rates than the ones 
defined in (\ref{rate2}):
\beq\label{sg1}\barr{lll}
\la_n=(3n+3c+1)(3n+3c+2)^2,\qq & n\geq 0 & \\[4mm]\dst 
\mu_n=(3n+3c)^2(3n+3c+1),\qq & n\geq 1 &\qq \mu_0\geq 0,\earr\eeq
where $\mu_0$ is taken as a free parameter, not necessarily equal 
to $(3c+1)(3c)^2.$

We will denote by $G_n(z;c,\mu_0)$ the polynomials with recurrence relation 
(\ref{op1}) and the rates (\ref{sg1}). Obviously the polynomials corresponding to 
process $P2$ are recovered as $G_n(z;c,0).$

In order to avoid repetitions, we will give only the main steps. 
It is again useful to define a triplet of polynomials $e_{3n+l}(\z),\ l=0,1,2$  
by the recurrence relation
\beq\label{sg2}
\z=z^{1/3}\qq n\geq 0 \qq 
\left\{\barr{l}
e_{3n+1}=-\z\,e_{3n}+\mu_n\,e_{3n-2},\\[4mm]  
e_{3n+2}=-\z\,e_{3n+1},\\[4mm] 
e_{3n+3}=-\z\,e_{3n+2}+\la_n\,e_{3n}, \earr\right.\eeq
with the boundary values
\[e_{-1}=-\frac 1{\z},\qq\qq e_0=1,\qq\Rightarrow\qq e_3=\la_0+\mu_0-z.\]
By an argument which follows closely the one given in the proof of Proposition 1, 
we get:

\begin{nth} The polynomials $G_n$ and $e_n$ are related by 
\beq\label{sg3}
G_n(z;c,\mu_0)=(3c+1)!\,\frac{(c+2/3)_n}{(c+1)_n}\,\frac{e_{3n}(\z)}{(3n+3c+1)!}.\eeq
\end{nth}

We then define the generating functions
\beq\label{sg4}
H_l(\z,t)=\sum_{n\geq 0}\,e_{3n+l}(\z)\,\frac{t^{3n+3c+l}}{(3n+3c+l)!},\qq l=0,1,2
\eeq
for which we get the differential system
\beq\label{sg5}
\barr{rcl} 
(1-t^3)\,D_t\,H_0-2t^2\,H_0+\z\,H_2 & = &\dst \frac{t^{3c-1}}{(3c-1)!}\\[5mm] 
D_t\,H_1+\z\,H_0 & = & 0\\[5mm]
(1-t^3)\,D_t\,H_2-t^2\,H_2+\z\,H_1 & = &\dst-\frac{\mu_0}{\z}\,\frac{t^{3c+1}}{(3c+1)!}
\earr\eeq
Switching from the functions $H_i$ to new functions $\wh{H}_i$ defined by 
\beq\label{sg6}
H_0=\frac{\wh{H}_0}{(1-t^3)^{2/3}},\qq G_1=\wh{H}_1,
\qq H_2=\frac{\wh{H}_2}{(1-t^3)^{1/3}},\eeq
and using the variable $\tht(t)$ defined in (\ref{fcts}) the previous system 
becomes an inhomogeneous differential system with constant coefficients, easy 
to solve.
Combining all this we get:

\begin{nth} The polynomials $G_n(z;c,\mu_0)$ have the generating function
\beq\label{sgtop1}\barr{l}\dst 
\sum_{n\geq 0}\frac{(c+1)_n}{(c+2/3)_n}\,G_n(z;c,\mu_0)\,t^{3n+3c+1}= 
3c(3c+1)\,\int_0^t\,\frac{\si_1(\z\Theta(t,u))}{\z}\,u^{3c-1}(1-u^3)^{-1/3}\,du
\\[6mm]\dst \hspace{6cm}
+\mu_0\,\int_0^t\frac{\si_2(\z\Theta(t,u))}{\z^2}\,u^{3c+1}(1-u^3)^{-2/3}\,du.
\earr\eeq
\end{nth}

In fact we need the generating function
\beq\label{gg}
{\cal G}(z;c,\mu_0)\equiv\sum_{n\geq 0}\,G_n(z;c,\mu_0).\eeq 
Let us prove:

\begin{nth} We have the relation
\beq\label{sgtop2}\barr{l}\dst
{\cal G}(z;c,\mu_0)=\frac{3}{B(c+2/3,1/3)} 
\left\{3c(3c+1)\int_0^1\,\frac{\si_2(\z\wh\tht(u))}{\z^2}
\,u^{3c-1}\,(1-u^3)^{-1/3}\,du\right.\\[6mm]\dst \left.\hspace{6cm}
+\mu_0\int_0^1\,\frac{1-\si_0(\z\wh\tht(u))}{z}\,u^{3c+1}(1-u^3)^{-2/3}\,du,\right\}
\earr\eeq
valid for $c>0$ and
\beq\label{sgtop3}\barr{l}\dst 
1-\frac z{\mu_0}\,{\cal G}(z;c,\mu_0)=\frac 3{B(c-1/3,1/3)}\,
\frac{(3c)^2(3c+1)}{\mu_0}\,\int_0^1\,u^{3c-2}(1-u^3)^{-2/3}\si_0(\z\wh\tht(u))\,du
\\[6mm]\dst 
+\frac 3{B(c+2/3,1/3)}\,\frac{\mu_0-(3c)^2(3c+1)}{\mu_0}\,\int_0^1\,u^{3c+1}
(1-u^3)^{-2/3}\si_0(\z\wh\tht(u))\,du,\earr\eeq
valid for $c>1/3.$
\end{nth}

\noindent{\bf Proof :}
In relation (\ref{sgtop1}) we change the variable $t$ to $\tau$ defined by 
$t=\tau^{1/3}$, then multiply both sides by $\tau^{-2/3}(1-\tau)^{-2/3}$ and 
integrate from $\tau=0$ to $\tau=1.$ The left hand-side is merely a Eulerian 
integral, while the right-hand side is a double integral. Interchanging the order 
of integrations, and using relations (\ref{b2}) one gets (\ref{sgtop2}). The 
integral which does not involve $\si$ functions can be expressed in terms of 
Euler Gamma functions. Then elemetary algebra yields \ref{sgtop3}).  $\qq\Box$

Equipped with these results, let us turn ourselves to the determination of the 
Nevanlinna matrix for the processes $P1$ and $P2.$

\section{The Nevanlinna matrices}
We will write the first Nevanlinna matrix as
\beq\label{nm1}
{\cal N}_1(z)=\left(\barr{cc} A_1(z) & C_1(z)\\[4mm]
B_1(z) & D_1(z)\earr\right).\eeq
As shown in \cite{Va1} one gets simpler results by considering the modified 
Nevanlinna matrix
\beq\label{nm1bis}
\wti{\cal N}_1(z)=\left(\barr{cc} \wti{A}_1(z) & C_1(z)\\[4mm]
\wti{B}_1(z) & D_1(z)\earr\right),\eeq
where
\[\wti{A}_1(z)\equiv A_1(z)-\frac{C_1(z)}{\alf},\qq 
\wti{B}_1(z)\equiv B_1(z)-\frac{D_1(z)}{\alf},\qq\quad 
-\frac 1{\alf}=\sum_{n\geq 1}\frac 1{\mu_n\pi_n}.\]

Let us begin with the computation of the modified Nevanlinna matrix 
for the process $P1.$

\subsection{The Nevanlinna matrix for P1}
We have first

\begin{nth}\label{1NN1} The modified Nevanlinna matrix $\wti{\cal N}_1(z)$ of 
process P1 can be expressed in terms of the generating functions 
${\cal F}$ and ${\cal G}$ as
\[\barr{ll}\dst 
\wti{A}_1(z)=\frac 1{\la_0}\,{\cal G}(z;c+1/3,\la_0)
-\frac 1{\la_1}\,{\cal G}(z;c+4/3,\la_1),                                
&\dst C_1(z)=1-\frac z{\mu_1}\,{\cal F}(z;c+1,\mu_1),\\[4mm]\dst 
\wti{B}_1(z)=-1+\frac z{\la_0}\,{\cal G}(z;c+1/3,\la_0), &\dst 
D_1(z)=z\,{\cal F}(z;c,0).\earr\]
\end{nth}

\noindent{\bf Proof :}

\noindent We use successively the relations proved in Lemma 6 of \cite{Va1}. 
For the reader's convenience we will recall these relations.
We have for the first element
\beq\label{nm3}
D_1(z)=z\sum_{n\geq 0}\,F_n(z;c,0)=z{\cal F}(z;c,0),\eeq
upon use of (\ref{ff}). The second element $C_1(z)$ is given by
\beq\label{nm4}
C_1(z)=1-\frac{z}{\mu^{(1)}_0}\sum_{n\geq 0}\,F^{(1)}_n(z;c),\eeq
where the polynomials $F^{(1)}_n(z;c)$ have the shifted rates 
\beq\label{nm5} 
\la^{(1)}_n=\la_{n+1},\qq\quad 
\mu^{(1)}_{n}=\mu_{n+1}\quad\Rightarrow\quad F_n^{(1)}(z;c)=F_n(z;c+1,\mu_1).\eeq
Using (\ref{ff}) we conclude to
\beq\label{nm6} 
C_1(z)=1-\frac z{\mu_1}{\cal F}(z;c+1,\mu_1).\eeq
The third element is given by
\beq\label{nm6bis}
\wti{B}_1(z)=-1+\frac z{\wti{\mu}_0}\sum_{n\geq 0}\wti{F}_n(z;c),\eeq
where the $\wti{F}_n(z;c)$ are the KMG duals of $F_n(z;c),$ given here by
\beq\label{nm7}
\wti{F}_n(z;c)=G_n(z;c+1/3,\la_0)\quad\Rightarrow\quad 
\wti{B}_1(z)=-1+\frac z{\la_0}{\cal G}(z;c+1/3,\la_0),
\eeq
where the last equality follows from (\ref{gg}). 
The last element is given by
\beq\label{nm7bis}
\wti{A}_1(z)=\frac 1{\wti{\mu}_0}\sum_{n\geq 0}\wti{F}_n(z;c)
-\frac 1{\wti{\mu}_1}\sum_{n\geq 0}\wti{F}_n^{(1)}(z;c).\eeq
We can write, using (\ref{gg})
\beq\label{nm8}
\wti{F}_n^{(1)}(z;c)=G_n(z;c+4/3,\la_1),\qq\quad \wti{\mu}_1=\la_1,\eeq
from which the proposition follows. $\qq\Box$

The Nevanlinna matrix follows quite easily now:

\begin{nth}\label{1NN}
The Nevanlinna matrix of the process P1, with rates
\[\la_n=(3n+3c+1)^2(3n+3c+2),\qq\mu_n=(3n+3c-1)(3n+3c)^2(1-\de_{n0}),\]
with $c>0,$ is given by
\beq\label{N1}\barr{l}\dst
\wti{A}_1=\frac 3{B(c+1,1/3)}\,\frac 1{(3c+1)}\,\int_0^1\,u^{3c}(1-u^3)^{-1/3}\ 
\frac{\si_2(\z\wh{\tht}(u))}{\z^2}\,du,\\[6mm]\dst 
\wti{B}_1=-\frac{3}{B(c+1,1/3)}\,\frac{3c}{3c+1}\,\int_0^1\,u^{3c-1}(1-u^3)^{-2/3}\ 
\si_0(\z\wh{\tht}(u))\,du,\\[6mm]\dst 
C_1=\frac 3{B(c+1/3,2/3)}\,\int_0^1\,u^{3c}(1-u^3)^{-1/3}\ 
\si_0(\z\wh{\tht}(u))\,du,\\[6mm]\dst 
D_1=\frac{3}{B(c+1/3,2/3)}\,3cz\,\,\int_0^1u^{3c-1}(1-u^3)^{-2/3}\
\frac{\si_1(\z\wh{\tht}(u))}{\z}\,du,\earr\eeq
where $\,\z=z^{1/3}.$
\end{nth}

\noindent{\bf Proof :}

\noindent The matrix element $D_1$ follows from Proposition \ref{1NN1} and 
(\ref{gftop2}). The matrix element $C_1$ follows from Proposition \ref{1NN1} and 
(\ref{gftop3}). The matrix element $\wti{B}_1$ follows from Proposition \ref{1NN1} 
and (\ref{gftop3}). To compute $\wti{A}_1$ we use Proposition \ref{1NN1} and 
(\ref{sgtop3}) to get first
\[z\,\wti{A}_1=-\frac{3}{B(c+1,1/3)}\,\frac 1{3c+1}\,\int_0^1\,\left[\rule{0mm}{5mm}
3cu^{3c-1}-(3c+1)u^{3c+2}\right](1-u^3)^{-2/3}\ \si_0(\z\wh{\tht}(u))\,du.\]
which is nothing but
\[z\,\wti{A}_1=-\frac{3}{B(c+1,1/3)}\,\frac 1{3c+1}\,\int_0^1\,D_u\left[\rule{0mm}{5mm}
u^{3c}(1-u^3)^{1/3}\right]\ \si_0(\z\wh{\tht}(u))\,du.\]
An integration by parts gives the required result in Proposition \ref{1NN}. $\qq\Box$

Let us observe that the matrix elements $D_1$ and $\wti{B}_1$ can be 
simplified for $c>1/3.$ Using relations (\ref{b2}) one realizes that an integration 
by parts is then possible, leaving us with the matrix elements
\beq\label{1NNsimp}\barr{l}\dst 
\wti{B}_1=-\frac{3}{B(c+1,1/3)}\,\frac 1{3c+1}\,\int_0^1\,u^{3c}\ 
\frac{\si_1(\z\wh{\tht}(u))}{\z}\,du,\\[6mm]\dst 
 D_1=\frac{3}{B(c+1/3,2/3)}\,\int_0^1u^{3c}\ \z\,
\si_2(\z\wh{\tht}(u))\,du.\earr\eeq
The special cases where $c=0$ and $c=1/3$ lead to considerable simplifications for 
$B$ and $D$ given in \cite{glv}.

\subsection{The Nevanlinna matrix for P2}
There is no need to give again the detailed proofs, since everything proceeds as 
for process $P1.$ Beware that now
\[\la_n=(3n+3c+1)(3n+3c+2)^2,\qq\mu_n=(3n+3c)^2(3n+3c+1)(1-\de_{n0}).\]
We have first

\begin{nth}\label{2NN1} The modified Nevanlinna matrix $\wti{\cal N}_2(z)$ of 
process P2 can be expressed in terms of the generating functions 
${\cal F}$ and ${\cal G}$ as
\[\barr{ll}\dst 
\wti{A}_2(z)=\frac 1{\la_0}\,{\cal F}(z;c+2/3,\la_0)
-\frac 1{\la_1}\,{\cal F}(z;c+5/3,\la_1),                                
&\dst C_2(z)=1-\frac z{\mu_1}\,{\cal G}(z;c+1,\mu_1),\\[4mm]\dst 
\wti{B}_2(z)=-1+\frac z{\la_0}\,{\cal F}(z;c+2/3,\la_0), &\dst 
D_2(z)=z\,{\cal G}(z;c,0).\earr\]
\end{nth}
Combining this result with the explicit forms of these generating functions, and 
upon integrations by parts, we get

\begin{nth}\label{2NN}
The Nevanlinna matrix of the process P2, with rates
\[\la_n=(3n+3c+1)(3n+3c+2)^2,\qq\mu_n=(3n+3c)^2(3n+3c+1)(1-\de_{n0}),\]
where $c>0,$ is given by
\beq\label{N2}\barr{l}\dst
\wti{A}_2=\frac 3{B(c+1,2/3)}\,\frac 1{(3c+2)}\,\int_0^1\,u^{3c}\ 
\frac{\si_2(\z\wh{\tht}(u))}{\z^2}\,du,\\[6mm]\dst 
\wti{B}_2=-\frac{3}{B(c+1,2/3)}\,\frac{3c}{3c+2}\,\int_0^1\,u^{3c-1}(1-u^3)^{-1/3}\ 
\si_0(\z\wh{\tht}(u))\,du,\\[6mm]\dst 
C_2=\frac 3{B(c+2/3,1/3)}\,(3c+1)\,\int_0^1\,u^{3c}\ 
\frac{\si_1(\z\wh{\tht}(u))}{\z}\,du,\\[6mm]\dst 
D_2=\frac{3}{B(c+2/3,1/3)}\,3c(3c+1)z\,\int_0^1u^{3c-1}(1-u^3)^{-1/3}\
\frac{\si_2(\z\wh{\tht}(u))}{\z^2}\,du,\earr\eeq
with $\,\z=z^{1/3}.$
\end{nth}
Here too, the $c\to 0$ limit gives again simplifications of the matrix
elements $B$ and $D$, see \cite{glv}.

We will now examine the asymptotics  of the entire functions appearing in the 
Nevanlinna matrix.

\section{Asymptotics of the Nevanlinna matrices}
Three quantities are essential to describe the large $|z|$ behaviour of an 
entire function $A(z)$, with Taylor series
\beq\label{entire}
A(z)=\sum_{n\geq 0}\xi_n(A) z^n.\eeq
The order $\rho_A$ is defined as
\beq\label{or1}
\rho_A=\limsup_{n\to\nf}\,\frac{n\ln n}{|\ln|\xi_n(A)||}.\eeq
If it is finite, we can define the type $\si_A$ as
\beq\label{or2}
\si_A=\frac 1{e\rho_A}\ \limsup_{n\to\nf}\,n|\xi_n(A)|^{\rho_A/n},\eeq
the Phragm\'en-Lindel\" of indicator $h_A(\phi)$ being defined by
\beq\label{or3}
h_A(\phi)=\limsup_{r\to\nf}\ \frac{\ln|A(r e^{i\phi})|}{r^{\rho_A}}
,\qq\qq\phi\in[0,2\pi].\eeq
The knowledge of the indicator gives the type via the relation
\beq\label{or5}
\si_A=\sup_{\phi\in[0,2\pi]}\ h_A(\phi).\eeq

As a preliminary remark, let us observe that all the matrix elements of the 
Nevanlinna matrices have the generic structure
\beq\label{struc1}
N_l(z)=\int_0^1\,f(u) E_l(z,u)\,du, \qq f(u)=u^a(1-u^3)^b,
\qq l=0,1,2,\eeq
possibly up to a single factor of $z,$ appearing in $D_1$ and $D_2.$ We will 
not care about this factor since it does not change the order, type and 
Phragm\'en-Lindel\" of indicator.

The entire functions $E_l$ which appear have the structure
\beq\label{struc2}
E_l(z,u)=\frac{\si_l(\z\wh{\tht}(u))}{\z^l}=\sum_{n\geq 0}
\frac{(-1)^n}{(3n+l)!}\wh{\tht}\,^{3n+l}(u)z^n.\eeq
By inspection, we see that the possible values of the parameters are 
\[a=3c,\  3c-1,\qq b=0,\ -1/3,\ -2/3\]
so that in any case we have
\beq\label{param}
a>-1\ \ \mbox{and}\ \ b\geq -2/3,\eeq
and these conditions ensure that $f$ is integrable over $[0,1].$ \\[0.2cm]

\noindent Let us begin with

\begin{nth}\label{order}
The order of the entire functions $N_l(z),\ l=0,\,1,\,2$ is $\rho=1/3.$
\end{nth}

\noindent{\bf Proof :}

\noindent  Since $E_l(z,u)$ are entire functions of $z$, uniformly in $u\in[0,1],$ 
we can integrate term by term in relation (\ref{struc2}). This gives
\beq\label{or7}
N_l(z)=\sum_{n\geq 0} \xi_{l,n}z^n,\qq\quad 
\xi_{l,n}=\frac{(-1)^n}{(3n+l)!}\int_0^1\,f(u)\wh{\tht}\,^{3n+l}(u)\,du.\eeq
From (\ref{or7}) we have
\beq\label{or8}
-\frac{\ln|\xi_{l,n}|}{n\ln n}=\frac{\ln\,(3n+l)!}{n\ln n}-
\frac{\ln\,I_{l,n}}{n\ln n},\qq I_{l,n}=\int_0^1\,f(u)\wh{\tht}\,^{3n+l}(u)\,du.
\eeq
In the large $n$ limit, using Stirling formula, this relation yields
\[\frac 1{\rho}=3-\limsup_{n\to\nf}\,\frac{\ln\,I_{l,n}}{n\ln n}.\]
To prove that the second term vanishes in that limit, (because of the logarithm)  
we need an upper and a lower bound for the integral. Using the inequalities 
given in (\ref{encadre}) we get
\[\tht_0^{3n+l}\int_0^1\,u^a(1-u)^{b+l+3n}\,du\leq I_{l,n}\leq \tht_0^{3n+l}\,M,\qq 
M=\int_0^1 f(u)\,du.\]
It follows that the logarithm of the upper and lower bounds behave, for large $n,$ 
respectively as 
\[(3n+l)\ln\tht_0-(a+1)\ln n,\qq\quad (3n+l)\ln\tht_0+\ln M,\]
and these imply
\[\limsup_{n\to\nf}\,\frac{\ln\, I_{l,n}}{n\ln n}=0,\]
and $1/\rho=3.\quad\Box$

It follows that all the matrix elements of the two Nevanlinna matrices have 
the same order $1/3.$ Let us state this result as:

\begin{nth}
All the matrix elements of the Nevanlinna 
matrices given by Proposition \ref{1NN} and Proposition \ref{2NN} have one and the 
same order $\rho=1/3.$
\end{nth}

Let us now determine the Phragm\'en-Lindel\" of indicator. We first need to prove 
the following lemma

\begin{nlem}Under the hypotheses (\ref{param}) we have, for $t\to +\nf$, the 
asymptotic behaviour
\beq\label{asy1}
I(t)\equiv\int_0^1\,u^a(1-u^3)^b\,e^{-tz\tht(u)}\,du=\frac{\G(a+1)}{t^{a+1}}
e^{-i(a+1)\phi}+{\cal O}\left(\frac 1{t^{a+4}}\right),\qq z=e^{i\phi},
\eeq
provided that $\phi\in]-\pi/2,+\pi/2[.$
\end{nlem}

\noindent{\bf Proof :}

\noindent The correspondence $u\to x=\tht(u)$ is a continuous bijection from $
[0,1]\to [0,\tht_0].$ It follows that its inverse function $\tht^{-1}$ is also a 
continuous bijection. Taking $x$ as a new variable, the integral $I(t)$ can be 
written as
\beq\label{asy2}
I(t)=\int_0^{\tht_0}\,\chi(x)\,e^{-tz x}\,dx,\qq
\chi(x)=[\tht^{-1}(x)]^a[1-(\tht^{-1}(x))^3]^{b+2/3}.\eeq
In view of the hypotheses (\ref{param}) the parameter $b+2/3$ is positive and 
$\chi(x)$ is 
continuous over $]0,\tht_0].$ Let us split the integral into two pieces:
\[I(t)=I_1(t)+I_2(t),\qq I_1(t)=\int_0^{\eps}\,\chi(x)\,e^{-tz x}\,dx,\qq
I_2(t)=\int_{\eps}^{\tht_0}\,\chi(x)\,e^{-tz x}\,dx,\] 
Since $\chi$ is continuous over $[\eps,\tht_0],$ we have
\[|I_2(t)|\leq \sup_{x\in[\eps,\tht_0]}\chi(x)\,
\frac{e^{-t({\rm Re}\,z)\eps}-e^{-t({\rm Re}\,z)\tht_0}}{t({\rm Re}\,z)},\]
and these terms are vanishing exponentially for large $t$, so they will be 
negligible when compared to inverse powers of $t.$
 
Next let us consider $I_1.$ For sufficiently small $\eps$ we can write
$\ \chi(x)=x^a+{\cal O}(x^{a+3}).$ So we get
\beq\label{inter1}
I_1(t)=\int_0^{\eps}\,[x^a+{\cal O}(x^{a+3})]e^{-tzx}\,dx=\int_0^{\nf}\cdots
-\int_{\eps}^{\nf}\cdots.\eeq
The second integral in (\ref{inter1}) is again exponentially small for large 
$t$ since we have
\[\left|\int_{\eps}^{\nf}\,x^a\,e^{-tzx}\,dx\right|\leq 
\int_{\eps}^{\nf}\,x^a\,e^{-t({\rm Re}\,z)x}\,dx\leq 
e^{-\eps t({\rm Re}\,z)/2}\int_{\eps}^{\nf}\,x^a\,e^{-t({\rm Re}\,z)x/2}\,dx,\]
and similarly for the term containing ${\cal O}(x^{a+3}).$ 
The further bounds
\[\left|\int_{\eps}^{\nf}\,x^a\,e^{-tzx}\,dx\right|\leq 
e^{-\eps t({\rm Re}\,z)/2}\int_0^{\nf}\,x^a\,e^{-t({\rm Re}\,z)x/2}\,dx=
\frac{\G(a+1)}{[t({\rm Re}\,z)/2]^{a+1}}e^{-\eps t({\rm Re}\,z)/2},\]
show that this term is also with an exponential decrease. 
The first integral in (\ref{inter1}) is well known to give
\[\int_0^{\nf}\,x^a\,e^{-tzx}\,dx=\frac{\G(a+1)}{t^{a+1}}e^{-i(a+1)\phi},\qq
z=e^{i\phi},\qq \phi\in ]-\frac{\pi}{2},+\frac{\pi}{2}[.\]
So we end up with \footnote{All the equalities are understood up to 
exponentially small terms omitted.}
\[I_1(t)=\frac{\G(a+1)}{t^{a+1}}e^{-i(a+1)\phi}+
\int_0^{\eps}\,{\cal O}(x^{a+3})e^{-tzx}\,dx.\]
For the last piece its modulus is  bounded by
\[\int_0^{\eps}\,x^{a+3}e^{-t({\rm Re}\,z)x}\,dx\leq
\int_0^{\nf}\,x^{a+3}e^{-t({\rm Re}\,z)x}\,dx={\cal O}\left(\frac 1{t^{a+4}}\right),\]
so the lemma is proved. $\quad\Box.$

In order to get the indicator let us point out that in the Nevanlinna matrix elements, 
only the entire functions $\si_l(\z\wh{\tht}(u))/\z^l,\ l=0,\,1,\,2$ do appear. Since 
the factor $\z^l$ does not change the indicator (see the definition (\ref{or3})), it 
is sufficient to deal with the 
$\si_l$ and from relation (\ref{sigma}) these can be written as linear combinations 
of exponentials. Considering a generic matrix element of the form
\beq\label{mat1}
N(z)=\int_0^1\,f(u)[a_0 e^{-\z\wh{\tht}(u)}+a_+ e^{j\z\wh{\tht}(u)}
+a_- e^{\ol{j}\z\wh{\tht}(u)}],\qq\z=r^{1/3}e^{i\phi/3},\qq z=\z^3,\eeq
we will prove:

\begin{nth}\label{PL}
All the functions $N$ have as Phragm\'en-Lindel\" of indicator 
\beq\label{PL3}
h(\phi)=\tht_0\,\cos\left(\frac{\phi-\pi}{3}\right),\qq\qq\phi\in[0,2\pi].\eeq
\end{nth}

\noindent{\bf Proof :}

\noindent Let us use the notations
\[z_0\equiv e^{i\phi/3}=c_0+is_0,\qq z_+\equiv e^{i(\phi+\pi)/3}=c_++is_+,
\qq z_-\equiv e^{i(\phi-\pi)/3}=c_-+is_-,\]
let us determine the Phragm\'en-Lindel\" of indicator first for 
$\phi\in]0,2\pi[.$ 

It is elementary to check that the following inequalities hold:
\beq\label{bornes}
\barr{lll}\dst \phi\in\left]0,\frac{\pi}{2}\right] &\qq c_->c_+>0 &\qq -c_0<0\\[5mm]
\dst \phi\in \left]\frac{\pi}{2},\frac{3\pi}{2}\right[ &\qq c_->\sqrt{3}/2 & -c_0<0,
\quad c_+<0\\[5mm]
\dst \phi\in\left[\frac{3\pi}{2},2\pi\right[ &\qq c_->-c_0>0 & \qq c_+<0\earr\eeq
The asymptotic behaviour of the integral must be analyzed separately for 
these three cases.

Let us begin with the simplest case, when
$\phi\in ]\frac{\pi}{2},\frac{3\pi}{2}[.$
One has
\[N(z)=e^{z_-r^{1/3}\tht_0}J(r,\tht)\]
with
\[J(r,\tht)=a_-\int_0^1\,f(u)e^{-z_-r^{1/3}\tht(u)}\,du
+e^{-z_-r^{1/3}\tht_0}\int_0^1\,f(u)[a_+e^{z_+r^{1/3}\wh{\tht}(u)}
+a_0e^{-z_0r^{1/3}\wh{\tht}(u)}]\,du.\]
We have merely used the relation $\wh{\tht}(u)=\tht_0-\tht(u)$ in the piece 
involving $a_-.$  
Since the real parts of $z_+$ and $-z_0$ are negative, the absolute value 
of the last two integrals 
is bounded. Furthermore, since the real part of $z_-$ is positive, for large $r$ 
the factor $e^{-z_-r^{1/3}\tht_0}$ has an 
exponentially decreasing absolute value. So we conclude to
\beq\label{detail1}
\limsup_{r\to\nf}\frac{\ln|N(re^{i\phi})|}{r^{1/3}}=c_-\,\tht_0+
\limsup_{r\to\nf}\frac{\ln\left|a_-\int_0^1\,f(u)e^{-z_-r^{1/3}\tht(u)}\,du\right|}
{r^{1/3}}.\eeq
The integral appearing in the numerator has a leading behaviour given by Lemma 1 
\[\left|\int_0^1\,f(u)e^{-z_-r^{1/3}\tht(u)}\,du\right|
\sim \frac{\G(a+1)}{r^{(a+1)/3}},\qq r\to\nf,\]
the hypothesis $(\phi-\pi)/3\in\,]-\pi/2,\pi/2[$ being indeed satisfied. This 
shows that the last term in the right hand side of (\ref{detail1}) vanishes; 
therefore we have obtained:
\beq\label{detail2}
h(\phi)=
\limsup_{r\to\nf}\frac{|N(re^{i\phi})|}{r^{1/3}}=
\tht_0\,\cos\left(\frac{\phi-\pi}{3}\right),\qq\quad
\phi\in \left]\frac{\pi}{2},\frac{3\pi}{2}\right[.\eeq

Let us now examine the region
$\,\phi\in ]0,\frac{\pi}{2}[.$ This time
$-c_0<0$ but $c_+>0$ remains still bounded by $c_-.$ So using again the relation 
$\wh{\tht}(u)=\tht_0-\tht(u)$ in the term involving $a_-$ and $a_+$ we can 
write this time
\[N(z)=e^{z_-r^{1/3}\tht_0}J(r,\tht)\]
with
\[\barr{l} J(r,\tht)=a_-\int_0^1\,f(u)e^{-z_-r^{1/3}\tht(u)}\,du
+a_+e^{-(z_--z_+)r^{1/3}\tht_0}\int_0^1\,f(u)e^{-z_+r^{1/3}\tht(u)}\,du\\[4mm]
\hspace{6.5cm}+a_0e^{-z_-r^{1/3}\tht_0}\int_0^1\,f(u)e^{-z_0r^{1/3}\wh{\tht}(u)}\,du.
\earr\]
The absolute value of the last two integrals vanishes exponentially for large $r$ so 
we have again the relations (\ref{detail1}), (\ref{detail2}) from which 
we conclude similarly that
\[h(\phi)=\tht_0\,\cos\left(\frac{\phi-\pi}{3}\right),\qq\quad
\phi\in \left]0,\frac{\pi}{2}\right].\]

The analysis for $\,\phi\in[\frac{3\pi}{2},2\pi[\,$ is similar to the one of the
previous interval, up to an interchange of $-c_0$ and $c_+.$ So we have obtained
the Phragm\'en-Lindel\" of indicator for $\,\phi\in]0,2\pi[.$ We then extend our
result to $[0,2\pi]$ using its continuity property and this ends the proof.
$\qq\Box$

Proposition \ref{PL} proves that all the Nevanlinna matrix elements have the 
same indicator, which we state as

\begin{nth}The Nevanlinna matrices given by Propositions \ref{1NN} and \ref{2NN}
have one and the same Phragm\'en-Lindel\" of indicator
\[h(\phi)=\tht_0\,\cos\left(\frac{\phi-\pi}{3}\right),\qq\quad
\phi\in [0,2\pi].\]
It follows from relation (\ref{or5}) that they have one and the same type
$\si=\tht_0.$
\end{nth}

\noindent{\bf Remarks :}

\brm
\item Our results agree with the general theorems proved in
\cite{BP} according to which, for a given Nevanlinna matrix, all of its
four elements have the same order, type and Phragm\'en-Lindel\" of indicator.
These quantities are therefore intrinsic to a given matrix.
Our Propositions \ref{order} and \ref{PL} allow us to deal, in one stroke,
with the two different Nevanlinna matrices obtained.
\item It may be interesting to observe that the quantities characterizing the
asymptotics in the complex plane of the Nevanlinna matrix are independent of
the parameter $c.$
\item It was shown in \cite{glv}, that for $c=0,$ the leading term in the asymptotic
behaviour of the N-extremal mass points was given by
\[x_n=\left(\frac{2\pi n}{\sqrt{3}\tht_0}\right)^3+o(n^3),\qq n\to\nf.\]
Since the order is always $1/3$ no matter what $c$ is, the leading cubic dependence
on $n$ will remain, but there is still an interesting open issue: will the
coefficient in front of $n^3$ remain independent of $c\ $?
\erm


\begin{thebibliography}{1111}
\bibitem{Ak} N. I. Akhiezer, {\sl The classical moment problem}, Oliver and Boyd, 
Edinburgh (1965).
\bibitem{BP} C. Berg and H. L. Pedersen, ``On the order and type of the entire 
functions associated with an indeterminate Hamburger moment problem", 
{\sl Arkiv. Math.}, {\bf 32} (1994) 1-11.
\bibitem{BV} C. Berg and G. Valent, ``The Nevanlinna parametrization for some 
indeterminate Stieltjes moment problems associated with birth and death processes", 
{\sl Methods Appl. Anal.}, {\bf 1} (1994) 169-209.
\bibitem{glv} J. Gilewicz, E. Leopold and G. Valent, ``New Nevanlinna matrices for
orthogonal polynomials related to cubic birth and death processes", Communication
at the 7th International Symposium on Orthogonal Polynomials, Copenhagen (2003),
to appear.
\bibitem{kmg} S. Karlin and J. L. McGregor,``The differential equations of 
birth-and-death processes, and the Stieltjes moment problem", {\sl Trans. Amer. 
Math. Soc.}, {\bf 85} (1958) 489-546.
\bibitem{Va1} G. Valent, ``Co-recursivity and Karlin-McGregor duality for 
indeterminate moment problems", {\sl Constr. Approx.}, {\bf 12} (1996) 531-553.

\end{thebibliography}
\end{document}